\documentclass[journal=jctcce,
]{achemso}
\usepackage{amsfonts}
\usepackage{amsmath}
\usepackage{amssymb}
\usepackage{graphicx}
\usepackage{dcolumn}
\usepackage{bm}
\usepackage{pdfpages}
\usepackage{float}
\usepackage{xcolor}

\setkeys{acs}{chaptertitle = true}

\newcommand{\func}[1]{\operatorname{#1}}
\title{Simulating Molecular Single Vibronic Level Fluorescence Spectra with ab initio Hagedorn Wavepacket Dynamics}
\author{Zhan Tong Zhang}
\author{Ji\v{r}\'i J. L. Van\'i\v{c}ek}
\email{jiri.vanicek@epfl.ch}
\affiliation{Laboratory of Theoretical Physical Chemistry, Institut des Sciences et
Ing\'enierie Chimiques, Ecole Polytechnique F\'ed\'erale de Lausanne (EPFL),
CH-1015 Lausanne, Switzerland}
\date{\today}

\begin{document}

\begin{abstract}

We present a practical, ab initio time-dependent method using Hagedorn wavepackets to efficiently simulate single vibronic level (SVL) fluorescence spectra of polyatomic molecules from arbitrary initial vibrational levels. 
We apply the method to compute SVL spectra of anthracene by performing wavepacket dynamics on a 66-dimensional harmonic potential energy surface constructed from density functional theory calculations. The Hagedorn approach captures both mode distortion (frequency changes) and mode mixing (Duschinsky rotation) within the harmonic approximation. We not only reproduce the previously reported simulation results for singly excited $12^1$ and $\overline{11}^1$ levels, but are also able to compute SVL spectra from multiply excited levels in good agreement with experiments. Notably, all spectra were obtained from the same wavepacket trajectory without any additional propagation beyond what is required for the emission spectrum from the ground vibrational level of the electronically excited state.

\end{abstract}

\graphicspath{{./figures/}{./Figures/}{"C:/Users/GROUP
LCPT/Documents/Tong/SVL_anthracene/Figures/"}}

\section{Introduction}

Single vibronic level (SVL) spectroscopy measures the fluorescence decay of a system following a vibronic excitation to a specific vibrational level in the excited electronic state.
This spectroscopic tool has played an important role in investigations of excited-state relaxation pathways,\cite{Parmenter_Schuyler:1970,Felker_Zewail:1985,Lambert_Zewail:1984,Duca:2004} vibrational structures of electronic states,\cite{Quack_Stockburger:1972,Lambert_Zewail:1984,Yang_Laane:2006} and in the identification and characterization of rotamers and reactive intermediates.\cite{Hollas_BinHussein:1989,Newby_Zwier:2009,Smith_Clouthier:2022}
Computational methods based on ab initio electronic structure calculations have been developed and implemented to better understand these spectra and guide experiments.\cite{Chau_Wang:1998,Chau_Mok:2001,BarbuDebus_Guchhait:2006,Lee_Chau:2007,Tapavicza:2019,Tarroni_Clouthier:2020,Sunahori_Clouthier:2022}

Calculations of SVL spectra have mostly relied on the time-independent approach, where individual Franck--Condon factors are computed for each vibronic transition.\cite%
{Chau_Wang:1998,Chau_Mok:2001,Grimminger_Sears:2013,Sunahori_Clouthier:2022}
However, for larger polyatomic molecules, the task becomes challenging due to the large number of possible transitions and the computation of nonseparable overlap integrals.\cite{Carrington:2011,Borrelli_Peluso:2013,Meier_Rauhut:2015,Conte_Ceotto:2023} The time-dependent approach, which is more natural for low- or intermediate-resolution
spectra, avoids the need to prescreen transitions and can more efficiently accommodate
Duschinsky rotation (mode mixing),\cite{Baiardi_Barone:2013,Wehrle_Vanicek:2014,Tapavicza:2019}
anharmonicity,\cite{Wehrle_Vanicek:2015,Bonfanti_Pollak:2018,Koch_Burghardt:2019,Begusic_Vanicek:2022}
Herzberg-Teller,\cite{
Huh_Berger:2012,Borrelli_Peluso:2012,Baiardi_Barone:2013,Patoz_Vanicek:2018,Begusic_Vanicek:2018,Bonfanti_Pollak:2018,Koch_Burghardt:2019,Prlj_Vanicek:2020,Kundu_Makri:2022} and temperature effects\cite{Borrelli_Peluso:2011,Borrelli_Gelin:2016,Begusic_Vanicek:2020,Begusic_Vanicek:2021a,Gelin_Borrelli:2023}  in both linear and multidimensional\cite{Schubert_Engel:2011,Begusic_Vanicek:2020a,Begusic_Vanicek:2021,Gelin_Domcke:2022} vibronic spectra.

In the first practical implementation of the time-dependent approach to SVL spectra,\cite{Huh_Berger:2012} Tapavicza developed a generating-function-based method to simulate emissions from a singly excited state (i.e., with at most a single vibrational excitation in one mode only).\cite{Tapavicza:2019} Using a displaced, distorted, and Duschinsky-rotated global harmonic model, he successfully computed SVL spectra of singly excited anthracene in good agreement with the experimental results.\cite{Lambert_Zewail:1984}

In order to evaluate SVL spectra associated with emission from arbitrary vibronic levels, we recently proposed\cite{Zhang_Vanicek:2024a} another time-dependent approach, one based on Hagedorn functions.\cite{Hagedorn:1998,Faou_Lubich:2009,Lasser_Lubich:2020}
These functions in the form of a Gaussian multiplied by carefully constructed polynomials\cite{Lasser_Troppmann:2014,Ohsawa:2019,Lasser_Lubich:2020} are exact solutions of the time-dependent Schr\"{o}dinger equation with a quadratic potential and have drawn attention
in applied mathematics\cite{Kargol:1999,Hagedorn_Joye:2000,Faou_Lubich:2009,Gradinaru_Joye:2010,Gradinaru_Joye:2010a,book_Lubich:2008,Lasser_Troppmann:2014,Dietert_Troppmann:2017,Ohsawa:2018,Lasser_Lubich:2020}
because of their promising applications in physics and chemistry.\cite{Bourquin_Hagedorn:2012,Kieri_Karlsson:2012,Zhou:2014,Gradinaru_Rietmann:2021,Gradinaru_Rietmann:2024}

The recursive expressions we derived for their overlap integrals\cite{Vanicek_Zhang:2025} made it possible to apply Hagedorn wavepackets in computational spectroscopy.\cite{Zhang_Vanicek:2024a, Zhang_Vanicek:2025}
While Hagedorn functions, as a complete orthonormal basis, can represent initial states of arbitrary shape, their form is particularly well-suited for simulating SVL spectra.
Within the harmonic and Condon approximations, the SVL process may be represented by a single Hagedorn function at all times and provides a clear, uncomplicated first demonstration of the Hagedorn dynamics approach to vibronic spectroscopy.
In ref~\citenum{Zhang_Vanicek:2024a}, we validated the Hagedorn approach to compute SVL spectra against quantum split-operator results in two-dimensional harmonic model systems incorporating displacement, mode distortion (changes of vibrational frequencies in different electronic states), and mode mixing (Duschinsky rotation) effects.

Here, we describe how the approach from ref~\citenum{Zhang_Vanicek:2024a} extends to realistic polyatomic molecules within the global harmonic approximation, which provides a convenient starting point for studying larger molecular systems and can guide the selection of ab initio methods for further on-the-fly local harmonic studies that include anharmonic effects.\cite{Zhang_Vanicek:2025}
In order to implement ab initio Hagedorn wavepacket dynamics, we neglect ro-vibrational coupling and perform the vibrational dynamics in the normal-mode coordinates on a full-dimensional, global harmonic potential energy surface derived from electronic structure calculations.
As a test, we simulate the SVL spectra of anthracene from $\overline{11}^{j}$ and $12^{j}$ ($j=1,2$) levels, for which assigned experimental data are available.\cite{Lambert_Zewail:1984}  We also predict the SVL spectra from the $12^{1}6^{1}$, $12^{1}\overline{5}^{1}$, and $12^{1}5^{1}$ levels, which remain to be measured experimentally.
Whereas the singly excited cases were already successfully simulated by Tapavizca,\cite{Tapavicza:2019} our Hagedorn approach can treat higher and mixed excitations from a single semiclassical trajectory by computing overlaps between Hagedorn functions at a small additional cost that becomes negligible in on-the-fly local harmonic applications.

\section{Methodology}

\subsection{Time-dependent approach to spectroscopy}

The traditional ``sum-over-states" approach for evaluating vibronic spectra relies on computing the Franck--Condon overlaps between the initial and final vibrational states.
In displaced and distorted harmonic potentials, 
these overlap integrals can be evaluated exactly using analytical formulas,\cite{Islampour_Lin:1999,
Chang:2005} or, when Duschinsky effects are considered, through recursive procedures.\cite{Sharp_Rosenstock:1964,Santoro_Barone:2007,Cerezo_Santoro:2022}
While this time-independent approach provides an explicit contribution from each vibronic transition to the spectrum, computational cost becomes prohibitive in larger molecules if one does not restrict the number of final vibrational states included in the calculation (either heuristically or through an automated procedure implemented in packages such as FCclasses \cite{Santoro_Barone:2007, Cerezo_Santoro:2022}).

An alternative, time-dependent approach avoids the need to pre-select and compute individual transitions. The spectrum is instead evaluated as the Fourier transform of an appropriate wavepacket autocorrelation function
\begin{equation}
    C(t)= \langle \psi_{0} | \psi_{t} \rangle,
    \label{eq:ct}
\end{equation}
i.e., the overlap between the initial nuclear wavepacket $\psi_{0}$ and the wavepacket $\psi_{t}$ propagated to time $t$ on the final electronic surface. 
In the case of SVL fluorescence spectrum, 
the emission rate per unit frequency from a vibrational level $|\psi_{0}\rangle = |{K}\rangle$ of the excited electronic state $e$ is given by
\begin{equation}
\sigma_{\text{em}}(\omega) = \frac{4\omega^3}{3\pi \hbar c^3} |{\mu}%
_{ge}|^2 \func{Re} \int^\infty_{0} \overline{C(t)} \exp[it(\omega -
\omega_{e,K})] \,dt,
\label{eq:spec}
\end{equation}
where
$K\equiv (K_1,\dots,K_D)$ is a multi-index specifying the initial vibrational quantum numbers in the $D$ normal modes, $\mu_{ge}$ is the electronic transition dipole moment (a constant scalar within the Condon approximation), and $\hbar\omega_{e,K}$ is the energy of
the initial vibronic state.
The nuclear wavepacket $|\psi_{t}\rangle = \exp(-i\hat{H}_{g}t/\hbar)|\psi_{0}\rangle$ in eq~\ref{eq:ct} is propagated with the ground-state Hamiltonian
$\hat{H}_{g}$.

In contrast to the time-independent approach, the dynamics-based approach does not explicitly enumerate the final vibrational states in the vibronic spectrum and obtains all peaks at once. The convergence of the spectrum (in terms of the frequency range and resolution) is determined by the time step used for evaluating autocorrelation functions and the total time of propagation. Although the individual contributions of each transition are not directly available, the time-dependent approach is more natural and computationally straightforward when one is interested in low- to intermediate-resolution spectra. It also enables the treatment of more general initial states (e.g., thermal ensembles or non-Condon effects) and potential energy surfaces (e.g, Duschinsky rotation and anharmonicity).


\subsection{Representation of SVL initial states by Hagedorn functions}

For the conventional emission spectrum from the ground level $K=\mathbf{0}$, the initial vibrational state may be represented by a normalized $D$-dimensional Gaussian wavepacket
\begin{equation}
\varphi_{0} (q) = \frac{1}{(\pi \hbar)^{D / 4} \sqrt{\det (Q_{t})%
}}\exp \left[ \frac{i}{\hbar} \left( \frac{1}{2} x^{T} \cdot P_{t}
\cdot Q_{t}^{- 1} \cdot x + p_{t}^{T} \cdot x + S_{t} \right) \right]  \label{eq:tga}
\end{equation}
in Hagedorn's parametrization, where $x:= q - q_{t}$ is the shifted position, $q_{t}$ and $p_{t}$ are the position and momentum of the
center of the wavepacket, $S_{t}$ is the classical action, and $Q_{t}$ and $P_{t}$ are
complex-valued $D$-dimensional matrices that satisfy the so-called symplecticity
conditions\cite{book_Lubich:2008,Lasser_Lubich:2020} (see eqs 5 and 6 of ref~\citenum{Zhang_Vanicek:2024a}) and determine the width matrix $\mathrm{A}_{t} := P_{t}\cdot Q_{t}^{-1}$ of the Gaussian.
We can then apply Hagedorn's raising operator
\begin{equation}
A^{\dagger} := \frac{i}{\sqrt{2 \hbar}}
\left(P_{t}^{\dagger} \cdot (\hat{q} - q_{t}) - Q_{t}^{\dagger} \cdot (\hat{p} - p_t) \right)
\end{equation}
to the Gaussian wavepacket $\varphi_{0}$ to recursively generate an orthonormal family of Hagedorn functions
\begin{align}
\varphi_{K + \langle j \rangle} &= \frac{1}{\sqrt{K_j + 1}} A_j^{\dagger}
\varphi_K \label{eq:hgdf}
\end{align}
in the form of multivariate polynomials multiplied by a common Gaussian, with 
$\langle j \rangle = (\underbrace{0, \dots, 0}_{j-1}, 1, \underbrace{0, \dots, 0}_{D-j})$ being the $D$-dimensional unit vector along the $j$-th degree of freedom.\cite%
{Hagedorn:1980,Hagedorn:1998,book_Lubich:2008,Lasser_Lubich:2020} 

The initial vibrational state $|\psi_{0}\rangle = |{K}\rangle$ in the SVL emission process can be exactly represented by a single Hagedorn function $\varphi_{K}$ in the normal-mode coordinates.
In the harmonic approximation, normal-mode coordinates for a given electronic state diagonalize the Hessian at its equilibrium geometry, and the ground vibrational wavefunction is a Gaussian (eq \ref{eq:tga}) with a diagonal width matrix $\mathrm{A}_{0} = P_{0}\cdot Q_{0}^{-1}$.
With both $Q_0$ and $P_0$ diagonal, the associated multidimensional Hagedorn function is
a product of univariate functions, each of which is a Gaussian multiplied by a Hermite polynomial. As we will see in the next sections, this simple Hermite factorization is lost during the dynamics on a general harmonic surface.

\subsection{Hagedorn wavepacket dynamics}

Remarkably, Hagedorn wavepackets are exact solutions to the time-dependent Schr\"{o}dinger equation (TDSE) with a harmonic potential
\begin{equation}
V_{g}(q)=v_{0,g}+(q-q_{\text{ref},g})^{T}\cdot \kappa_{g} \cdot (q-q_{\text{ref},g})/2,
\label{eqn:harmonic_pot}
\end{equation}
which, in the calculations of emission spectra, represents the potential energy surface of the ground electronic state. The reference position $q_{\text{ref},g}$, the reference energy $v_{0,g}$, and the Hessian matrix $\kappa_{g}$ may be determined from ab initio electronic structure calculations.
The time evolution of a Hagedorn wavepacket then follows particularly simple, classical-like equations
\begin{align}
\dot{q_t} &= m^{-1} \cdot p_t, & \dot{p_t} &= -V_{g}^{\prime}(q_t)\nonumber = -\kappa_{g} \cdot (q_t-q_{\text{ref}}) \\ 
\dot{Q_t} &= m^{-1} \cdot P_t, & \dot{P_t} &= -V_{g}^{\prime\prime}(q_t) \cdot Q_t =-\kappa_{g} \cdot Q_t,\nonumber \\
\dot{S_t} &= L_t = p^{T}\cdot m^{-1}\cdot p/2-V_{g}(q_t),\label{eq:prop}
\end{align}
where $m$ is the mass matrix and $L_{t}$ is the Lagrangian.\cite{book_Lubich:2008,Lasser_Lubich:2020,Vanicek:2023}
The multi-index $K$ of the Hagedorn function does not change during the exact propagation in up-to-quadratic potentials, i.e., the Hagedorn function propagated in a harmonic potential to time $t$
retains its form 
\begin{equation}
	\varphi_{K,t} = (K!)^{-1 / 2}\left(A^{\dagger}[q_t,p_t,Q_t, P_t]\right)^K \varphi_0[q_t,p_t,Q_t, P_t,S_t],
\end{equation}
even in the presence of mode mixing (when Hessian $\kappa_{g}$ is not a diagonal matrix).
In fact, equations of motion (\ref{eq:prop}) are the same as in the thawed Gaussian approximation,\cite{Heller:1975,Heller:1976a}
which has been applied to vibronic spectra from ground vibrational levels\cite{Wehrle_Vanicek:2014,Wehrle_Vanicek:2015,Begusic_Vanicek:2019,Kletnieks_Vanicek:2023} and extended to propagate not only Gaussians but also Gaussians multiplied by a linear polynomial prefactor.\cite{Lee_Heller:1982,Patoz_Vanicek:2018,Begusic_Vanicek:2020,Prlj_Vanicek:2020,Wenzel_Mitric:2023,Wenzel_Mitric:2023a}
Using Hagedorn wavepackets, it is possible to treat arbitrary polynomials times a Gaussian (describing, e.g., excited vibrational states), and since the propagation (\ref{eq:prop}) still depends only on the Gaussian's parameters, a single Gaussian trajectory is sufficient to obtain SVL spectra from all initial vibrational levels.

Although equations (\ref{eq:prop}) can also be solved analytically to obtain a solution at an arbitrarily long time $t$ in a harmonic system,\cite{book_Tannor:2007} the spectrum (\ref{eq:spec}) depends on the values of the autocorrelation function $C(t)$ at all times $t$ from 0 to $t_{\text{max}}$ (the time that gives the desired spectral resolution). Thus, compared to evaluating analytical solutions at each discretized time point over the total time of propagation, numerical propagation (with geometric integrators) is nearly as accurate and efficient.
In high-dimensional harmonic systems, the computational cost of evaluating SVL spectra is dominated by computing the Hagedorn overlaps rather than the propagation, which only needs to be performed once for all initial states.

The numerical approach to wavepacket propagation becomes necessary when the Hagedorn dynamics is combined with the local harmonic approximation
to take into account anharmonicity in flexible molecules. There, as in the thawed Gaussian approximation, only minimal modifications need to be made to eqs \ref{eq:prop} by evaluating the gradient $V_{g}^{\prime}(q_t)$ and Hessian $V_{g}^{\prime\prime}(q_t)$ on the fly from electronic structure calculations (instead of from the static Hessian matrix $\kappa_{g}$ in the global harmonic model), but analytical solutions at arbitrary times are no longer possible. In the local harmonic case, the cost of evaluating SVL spectra is dominated by ab initio calculations and the cost of computing overlaps (in post-processing) becomes negligible.

Whereas closed-form analytical expressions exist for overlaps between two Gaussian wavepackets, computing the SVL spectra requires the overlaps between two ``excited" Hagedorn functions associated with two different Gaussian centers, a problem similar to the computation of Franck--Condon factors in harmonic systems in the time-independent approach. In ref~\citenum{Vanicek_Zhang:2025}, we derived recurrence relations for algebraically evaluating the overlaps between any Hagedorn functions; these expressions are also presented succinctly in the Supporting Information here.
 Although more complex than simple Gaussian overlaps, these calculations typically involve only one or two initially excited modes containing just a few vibrational quanta.

Beyond the $O(D^3)$ scaling with the number of normal modes $D$ (the same as in computing overlaps of Gaussian), the cost of evaluating the overlaps between Hagedorn functions depends on the number of excited modes and the vibrational excitation levels.  For multiply excited levels in a single mode, the cost of evaluating $\langle\varphi_{K,t=0}|\varphi_{K,t}\rangle$ scales quadratically with the excitation $K$. 

Compared to the sum-over-states approach, which requires overlaps between an initial vibrational excited state and a considerable number of possible final states in high-dimensional systems, the time-dependent Hagedorn method requires only overlaps between Hagedorn functions with the same multi-index $K$ (determined by the initial vibrational excitation).
The recursive nature of the overlap expressions also means that computing autocorrelation functions for multiply excited levels inherently generates those for lower levels encountered during recursion (e.g., computing the autocorrelation for a doubly excited level also provides results for singly excited levels). Conversely, with efficient caching implemented, results for lower excitation levels can be reused for higher excitation levels.

\subsection{Combination with ab initio electronic structure calculations}

The vibrational nuclear dynamics represented by Hagedorn wavepackets is described in terms of vibrational normal-mode coordinates, which naturally emerge from the harmonic approximation of the potential.
In order to make the Hagedorn approach practical for ab initio applications in real molecules, the initial state $\varphi_{K}$ and the potential energy surface (eq \ref{eqn:harmonic_pot}) must be constructed from electronic structure calculations. Because these calculations are typically carried out in Cartesian or internal coordinates, the computed molecular geometries and Hessians must be first transformed to normal-mode coordinates.
For simplicity of propagation, we adopt mass-weighted coordinates so that each normal mode has an identical mass and the mass matrix $m$ becomes scalar.

The first step in transforming Cartesian coordinates to mass-weighted normal-mode coordinates is minimizing the ro-vibrational coupling relative to a reference geometry. This is achieved by moving to the center-of-mass frame and applying the Kabsch algorithm to satisfy the translational and rotational Eckart conditions.\cite{Eckart:1935, Kudin_Dymarsky:2005, Kabsch:1978}
Next, the normal modes and the corresponding transformation matrix are found by diagonalizing the mass-weighted Hessian matrix of the reference state and projecting out the rotational and translational degrees of freedom (see section 6.7 of ref~\citenum{Vanicek_Begusic:2021}).
In the SVL application, the reference geometry and Hessian are taken from the initial, excited electronic state so that each vibrational eigenstate $|K\rangle$ may be represented by a single Hagedorn function $\varphi_K$.
The diagonalized vibrational Hessian $\kappa_e$ in the excited-state normal-mode coordinates determines the width matrix $\mathrm{A_0} = i\sqrt{m^{1/2}\cdot \kappa_e \cdot m^{1/2}}$ of the Gaussian $\varphi_{0}$ associated with the SVL initial states.
In Hagedorn's parametrization, we set $Q_0 = (\operatorname{Im} \mathrm{A}_0)^{-1/2}$ and $P_0 = \mathrm{A}_0 \cdot Q_0$ to satisfy the symplecticity relations.

In the adiabatic harmonic approximation, the Hessian $\kappa_{g}$ defining the ground-state potential (eq \ref{eqn:harmonic_pot}) used for propagation is evaluated at the ground-state equillibrium geometry $q_{\text{ref},g} = q_{\text{eq},g}$; both the geometry and Hessian are transformed to be in the excited-state mass-weighted normal-mode coordinates.
In the vertical harmonic approximation, the ground-state Hessian $\kappa_{g}$ is instead evaluated at the optimized excited-state geometry $q_{\text{ref},g} = q_{\text{eq},e}$. 
Due to mode mixing, $\kappa_{g}$ is in general not diagonal when expressed in the excited-state normal-mode coordinates.
Therefore, the initially diagonal $Q_{t=0}$ matrix of the SVL initial state will no longer be diagonal after evolution with the ground-state potential, and the propagated Hagedorn function will not remain a simple product of univariate Hermite functions.\cite{Lasser_Troppmann:2014,Ohsawa:2019,Lasser_Lubich:2020} However, 
 a single Hagedorn function still suffices to represent the SVL process exactly in the harmonic limit.
 In contrast, basis expansion methods (using, e.g., products of Gaussians and Hermite polynomials) typically require many functions to maintain accuracy in the presence of mode mixing in multi-dimensional systems.

\section{Computational details}

To compare with previously reported results and to validate our method, we used the same density functional theory (DFT) method (at PBE0/def2\nobreakdash-TZVP level of theory\cite{Adamo_Barone:1999,Weigend_Ahlrichs:2005}) as in ref~\citenum{Tapavicza:2019} to construct the global harmonic models. 
Geometry optimizations and frequency calculations were performed using the Gaussian 16 package\cite{software_g16} for the ground ($S_0\,\mathrm{^{1}A_{g}}$) and the first excited ($S_1\,\mathrm{^{1}B_{2u}}$) electronic states of anthracene. The excited-state calculations were carried out using standard linear-response time-dependent DFT.
While this level of theory is somewhat simple and approximate, it was shown\cite{Tapavicza:2019} to reproduce emission spectra from the ground and singly excited levels reasonably well with  results similar to those obtained from more expensive second-order approximate coupled-cluster calculations.
The optimized structures of the two states and the frequencies of the vibrational modes analyzed in this work are listed in the Supporting Information. 
The assignment of the vibrational modes and their symmetry follows the convention of ref~\citenum{Lambert_Zewail:1984} and the Supporting Information of ref~\citenum{Tapavicza:2019}.

A second-order TVT geometric integrator~\cite{book_Hairer_Wanner:2006,book_Lubich:2008,Vanicek:2023} was used to propagate the parameters of the Gaussian for a total time of $8\times 10^{5}\,\text{au}$ ($\sim$19\,ps) with a time step of 8\,au in the global harmonic systems considered in this work.
After the (single!) trajectory of the parameters of the Gaussian was obtained in a given harmonic system, the autocorrelation functions between Hagedorn wavepackets were computed every four steps for each initial vibrational level ($0^0, 12^1, 12^2, \overline{11}^1, \overline{11}^2, 12^{1} 6^{1}, 12^{1}\overline{5}^{1}$, and $12^{1}5^{1}$)
using the algebraic algorithm described in ref~\citenum{Vanicek_Zhang:2025} (a Python implementation is provided in the supplementary material of ref~\citenum{Zhang_Vanicek:2024a}).
Given the limited range of the observed experimental data, it was unnecessary to extend the range of the simulated spectra by computing the autocorrelation function at every step. A Gaussian damping function with a half-width at half-maximum of 20000\,au was applied to the autocorrelation functions before performing the Fourier transform.

To facilitate the comparison, the intensities of the highest peaks in all spectra (simulated and experimental) were set to unity.
A horizontal shift, determined by aligning the $0^{0}_{0}$ peak in the computed ground-level ($0^0$) fluorescence spectrum (see the Supporting Information) to be at the experimental origin $27709\,\text{cm}^{-1}$, was applied to the wavenumbers of the simulated spectra in global harmonic potentials to correct for the relatively large error (compared to the vibrational features) in the computed electronic excitation energy; the $\omega^3$ factor in eq \ref{eq:spec} was then adjusted based on the corrected wavenumbers (see the Supplemental Material of ref~\citenum{Zhang_Vanicek:2025} for details).

The details of the artificially modified harmonic systems (analyzed in fig~\ref{fig:11a_rot}) and of the local harmonic calculations (shown in fig~\ref{fig:lha_vha_spectra}) are available in the Supporting Information.

\section{Results and Discussion}

\begin{figure}[!ht]
    \centering
    \includegraphics[width=0.8\linewidth]{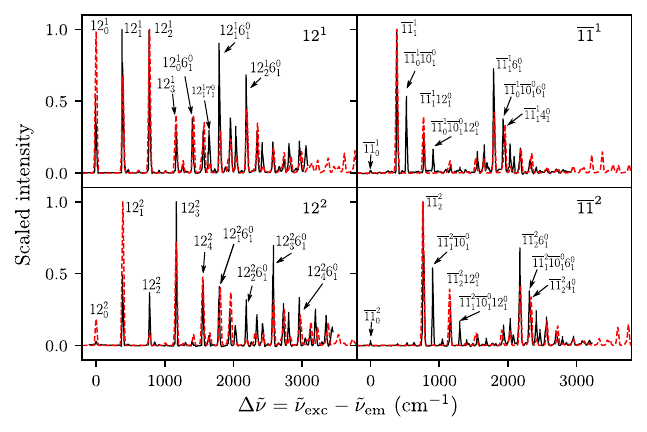}
    \caption{SVL fluorescence spectra of anthracene from initial vibrational levels $12^{j}$ and $\overline{11}^{j}$ ($j=1,2$) computed from Hagedorn wavepacket dynamics in the adiabatic harmonic model; a scaling factor of 0.97 was applied to the wavenumbers of the computed spectra (red dashed line); the
    experimental reference (black solid line) is taken from ref~\citenum{Lambert_Zewail:1984}.}
    \label{fig:spectra}
\end{figure}

Anthracene belongs to the $D_{2h}$ point group.
Within the Condon approximation,
only vibrational levels belonging to the $\mathrm{a_g}$ representation can be prepared in the excited $\mathrm{{}^{1}B_{2u}}$ state from the ground vibrational level of the ground $\mathrm{^{1}A_{g}}$ state to serve as the initial states in SVL experiments.\cite{Lambert_Zewail:1984,Tapavicza:2019}
However, through Herzberg--Teller intensity borrowing, the preparation of SVL initial states with $\mathrm{b_{1g}}$ vibrational symmetry is also possible.
Out of the 66 vibrational modes, twelve have $\mathrm{a_g}$ symmetry and eleven have $\mathrm{b_{1g}}$ symmetry (denoted by a bar in the mode label).
As examples, we show below the SVL spectra with initial vibrational excitations in the $\mathrm{a_g}$ mode 12 and $\mathrm{b_{1g}}$ mode $\overline{11}$.

Figure \ref{fig:spectra} compares the spectra of anthracene evaluated with the Hagedorn approach in the adiabatic harmonic model to the experimental emission spectra from the $12^1$, $12^2$, $\overline{11}^{1}$, and $\overline{11}^2$ levels.\cite{Lambert_Zewail:1984} 
For these SVL spectra, the Franck--Condon selection rules for emissions provide that transitions from $\mathrm{a_g}$ levels ($12^1$, $12^2$, and $\overline{11}^{2}$) are only allowed to $\mathrm{a_g}$ levels, and that transitions from the $\mathrm{b_{1g}}$ level ($\overline{11}^{1}$) are only allowed to $\mathrm{b_{1g}}$ levels. The total quantum numbers in $\mathrm{b_{1g}}$ modes must thus retain the same parity (odd or even) after the transition.

In fig~\ref{fig:spectra}, the computed and experimental SVL fluorescence spectra are shown with respect to the wavenumber differences $\tilde{\nu}_\text{exc} - \tilde{\nu}_\text{em}$ between the initial excitation light and the emitted light.
An empirical scaling factor was applied to the wavenumber differences of the simulated spectra since the density functional theory (DFT) calculations systematically overestimate the vibrational frequencies of the ground electronic state.\cite{Scott_Radom:1996,Tantirungrotechai_Limtrakul:2006,Tapavicza:2019}
By optimizing the alignment of peak positions between the simulated and experimental spectra, we set the scaling factor to 0.97, which is close to the vibrational scaling factors reported for the PBE0 functional with similar basis sets.\cite{nist_cccbdb:2022}

The simulated spectra from both singly excited levels (first row in fig \ref{fig:spectra}) agree well with the experiment and are consistent with the results obtained by Tapavicza using a generating function approach (see fig~S2 in the Supporting Information). Both Tapavicza's approach and our Hagedorn method are exact in harmonic models and are thus equivalent for singly excited levels. The overall structure of the $12^1$ spectrum is reproduced, including the progression of mode 12 ($12^{1}_{j}, j = 0,1,2,3$) and the transitions involving a mode change to another $\mathrm{a_g}$ mode (e.g., $12^{1}_{0}6^{0}_{1}$) or combinations with other $\mathrm{a_g}$ modes (e.g., $12^{1}_{1}6^{0}_{1}$ and $12^{1}_{2}6^{0}_{1}$).
While the intensity of the $12^1_0$ peak is overestimated, the computed intensities of certain combination bands (for example, $12^1_1 7^0_1$, $12^1_1 6^0_1$ and $12^1_2 6^0_1$) are significantly lower than experimentally observed.
From the  $\overline{11}^1$ level, the $\overline{11}^1_0$ transition is negligible in the experimental spectrum and entirely absent in the computed spectrum, as expected from symmetry considerations. Whereas the harmonic Hagedorn dynamics reproduces the cluster of combination bands between 1500 and 2500\,$\text{cm}^{-1}$, the peaks of $\overline{11}^{1}_{0}\overline{10}^{0}_{1}$, $\overline{11}^{1}_{0} \overline{10}_1^0 12_1^0$
, and $\overline{11}^{1}_{0} \overline{10}_1^0 6_1^0$
transitions are severely underestimated or missing in the computed spectrum.
The harmonic model allows clear peak assignments in the computed spectra. For example, while the $\overline{11}^{1}_{0} \overline{10}_1^0 6_1^0$ and $\overline{11}^{1}_{1} 4_1^0$ peaks are overlapping in the experiment, the peak present at 1954\,$\text{cm}^{-1}$ (2015\,$\text{cm}^{-1}$ before wavenumber scaling) in the computed $\overline{11}^1$ spectrum can be unambiguously assigned to the $\overline{11}^{1}_{1} 4_1^0$
instead of $\overline{11}^{1}_{0} \overline{10}_1^0 6_1^0$
transition based on the ab initio vibrational frequencies.

In contrast to the alternative method from ref~\citenum{Tapavicza:2019}, the Hagedorn approach makes it possible to compute SVL spectra from multiply excited levels and using the same trajectory as that already needed for the ground-level emission. 
The ab initio Hagedorn results for levels $12^{2}$ and $\overline{11}^{2}$ (second row in fig \ref{fig:spectra}) agree reasonably well with the experiments.
The computed $12^{2}$ spectrum correctly captures the experimentally observed decrease (compared to the $12^{1}$ spectrum) in the intensities of the $12^j_0$, $12^{j}_{j}$, and $12^{j}_{j}6^{0}_{1}$ peaks.

The experimental SVL spectra from $\overline{11}^{1}$ and $\overline{11}^{2}$ levels are broadly similar in structure.
Symmetry requires that the initial and final total vibrational quantum numbers in the $\mathrm{b_{1g}}$ modes maintain the same parity. However, a parity-allowed transition may still have little or no intensity.
For example, although allowed by symmetry, the $\overline{11}^{2}_{0}$ transition ($a_{\mathrm{g}}\rightarrow a_{\mathrm{g}}$) appears in neither the experimental nor the computed spectrum. Instead, the overall $\overline{11}^{2}$ spectrum is ``shifted" compared to the $\overline{11}^{1}$ spectrum, with the $\overline{11}^{2}_{2}$ transition serving as a false origin.
 The computed $\overline{11}^{2}$ spectrum, like the $\overline{11}^{1}$ spectrum, fails to describe certain combination peaks (e.g., $\overline{11}^2_1\overline{10}_1^0$ and $\overline{11}_1^2 \overline{10}_1^0 12_1^0$).

As demonstrated in ref~\citenum{Zhang_Vanicek:2024a}, the Hagedorn wavepacket dynamics is exact in harmonic potentials and can treat mode distortion (frequency changes) and mode mixing (Duschinsky rotation) exactly.
Therefore, the differences between experimental and simulated spectra must be due to the inaccuracies of the global harmonic model from electronic structure calculations, neglect of Herzberg--Teller contributions (particularly for $\mathrm{b_{1g}}$ levels), significant anharmonicity of the true potential energy surface, or possibly (but unlikely) experimental error.
To better understand the discrepancies, 
we show below the effects of different global harmonic approximations and the influence of Duschinsky coupling on the computed SVL spectra of anthracene. The importance of anharmonicity is also examined by comparing spectra from vertical, adiabatic, and local harmonic approximations.

\begin{figure}
    \centering
    \includegraphics[width=\linewidth]{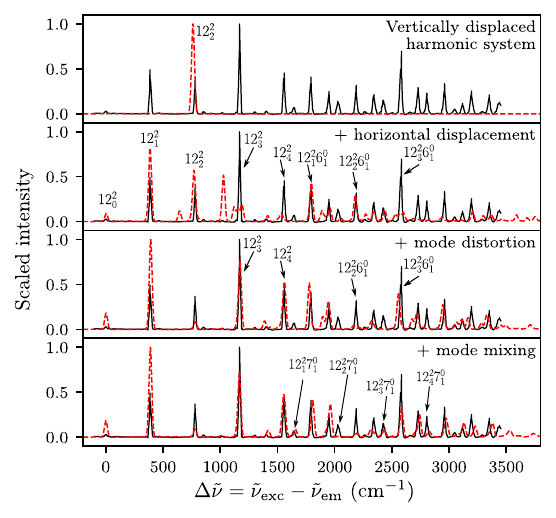}
    \caption{Effects of displacement, mode distortion, and mode mixing on the simulated (red dashed line) $12^{2}$ SVL fluorescence spectra of anthracene; a scaling factor of 0.97 was applied to the wavenumbers of the computed spectra. All spectra are compared to experiment (ref~\citenum{Lambert_Zewail:1984}, black solid line).}
    \label{fig:effects}
\end{figure}

Figure \ref{fig:effects} shows the effects of displacement, mode distortion, and mode mixing on the computed $12^{2}$ spectrum.
In the first row, we assumed the same harmonic potential energy surface for both excited- and ground-state surfaces (differing only by the adiabatic electronic energy gap).
The vibrational eigenfunctions with different vibrational quantum numbers are then orthogonal, and the simple selection rule in this vertically displaced harmonic system results in only a single peak (corresponding to the $12^{2}_{2}$ transition) in the spectrum.

The displacement of the equilibrium position of the ground-state surface from the excited-state equilibrium produces more interesting, nontrivial spectral features (second row of fig~\ref{fig:effects}). 
In the vertically and horizontally displaced harmonic oscillator model, the wavepacket was propagated with a potential centered around the ground-state equilibrium geometry but whose Hessian is identical to the excited-state Hessian (i.e., $\kappa_{g} = \kappa_{e}$ in $V_{g}$). Several experimental features (e.g., $12^{2}_{0}$, $12^{2}_{1}$, $12^{2}_{2}$, $12^{2}_{1}6^{0}_{1}$, and $12^{2}_{2}6^{0}_{1}$) are now reproduced, but some peaks (e.g., $12^{2}_{3}$, $12^{2}_{4}$, and $12^{2}_{3}6^{0}_{1}$) are underestimated or not resolved in the computed spectrum.
The description of the $12^{2}_{3}$, $12^{2}_{4}$, and $12^{2}_{3}6^{0}_{1}$ peaks is improved in the third row by including
mode distortion in the ground state (though the intensity of the $12^{2}_{2}6^{0}_{1}$ peak becomes worse).
Mode distortion, which allows frequency changes between the ground- and excited-state surfaces,
was included by defining the ground-state potential using only the diagonal components of the computed ground-state Hessian at the ground-state equillibrium geometry. As the off-diagonal Hessian elements were set to zero, the mode mixing effects were completely neglected in the computed spectrum shown in the third row.

Finally, the fourth row of fig~\ref{fig:effects} (the same as the bottom left panel of fig~\ref{fig:spectra}) shows the computed spectrum when the mode mixing is also taken into account. This most comprehensive harmonic description of the potential energy surface (as computed by DFT) improves agreement with experimental data, particularly for transitions with combination levels, for example, the $12^{2}_{j}7^{0}_{1}$ ($j=1,3,4$) peaks, although the intensities are not perfect. The $12^{2}_{2}7^{0}_{1}$ transition also appears with minimal intensity in the computed spectrum; its extremely low intensity, despite the inclusion of mode mixing, is partly because the $12^{2}_{2}$ transition is underestimated in our harmonic model.

\begin{figure}
    \centering
    \includegraphics[width=\linewidth]{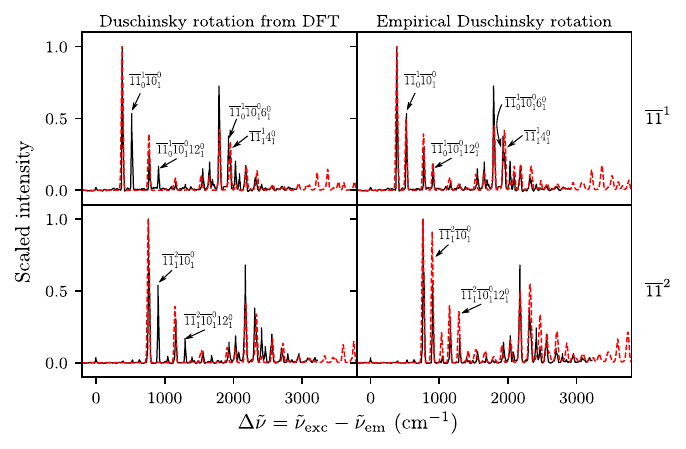}
    \caption{Effects of an artificially enhanced Duschinsky coupling between modes $\overline{11}$ and $\overline{10}$ on the computed $\overline{11}^{1}$ and $\overline{11}^{2}$  SVL fluorescence spectra of anthracene; a scaling factor of 0.97 was applied to the wavenumbers of the computed spectra (red dashed line); the
    experimental reference (black solid line) is taken from ref~\citenum{Lambert_Zewail:1984}.}
    \label{fig:11a_rot}
\end{figure}

The large observed effects of mode mixing on combination band intensities suggest that the tendency of our ab initio results to underestimate the intensities of certain combination transitions is due to an inaccurate description of Duschinsky coupling between the involved modes.
Indeed, the author of ref~\citenum{Tapavicza:2019} improved simulated SVL spectra by empirically fitting Duschinsky rotation matrices between several $a_{\textrm{g}}$ modes.
To further investigate the deficiencies of the $\overline{11}$ spectra computed with the ab initio harmonic Hagedorn dynamics,
 we applied an artificial Duschinsky rotation between modes $\overline{11}$ and $\overline{10}$ in the computed Hessian matrix (see the Supporting Information for details).
As shown in fig~\ref{fig:11a_rot}, the $\overline{11}^{1}$ spectrum (top right) evaluated with this modified potential recovers the transitions ($\overline{11}^{1}_{0}\overline{10}^{0}_{1}$, $\overline{11}^{1}_{0} \overline{10}_1^0 12_1^0$, and $\overline{11}^{1}_{0} \overline{10}_1^0 6_1^0$)
 absent in the original results (top left) and reproduces the experimental intensities of these peaks well. The previously missing peaks ($\overline{11}^2_1\overline{10}_1^0$ and $\overline{11}_1^2 \overline{10}_1^0 12_1^0$) in the computed $\overline{11}^{2}$ spectrum (bottom left) also appear when the artificial coupling is applied (bottom right), but the intensities of the recovered peaks are not as accurate.
The results confirm that underestimated mode coupling between vibrational modes is likely the primary cause of the missing peaks in the SVL spectra from $\overline{11}$ levels.

 \begin{figure}
    \centering
    \includegraphics[width=\linewidth]{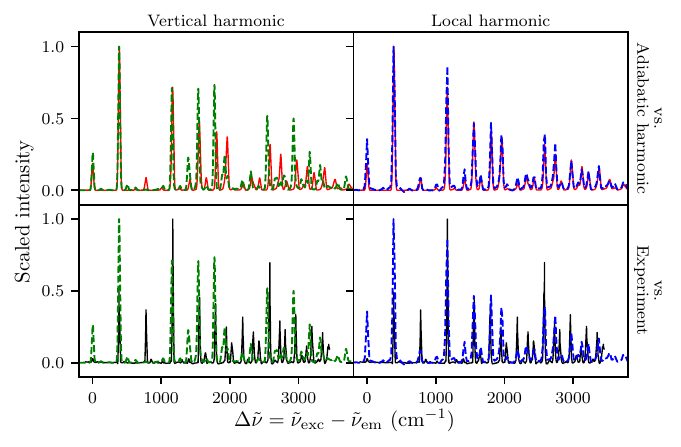}
    \caption{$12^{2}$ SVL fluorescence spectra of anthracene evaluated with the vertical (left, green dashed line) and local harmonic (right, blue dashed line) approaches. The spectra are compared to the adiabatic harmonic (first row, red solid line) and the experimental (ref~\citenum{Lambert_Zewail:1984}, second row, black solid line) spectra; a scaling factor of 0.97 was applied to all computed spectra.}
    \label{fig:lha_vha_spectra}
\end{figure}

The need for empirical wavenumber scaling in our results (see the Supporting Information for corresponding spectra without wavenumber scaling) indicates that anharmonicity may also be important. 
To test the adequacy of the global harmonic model, we first computed spectra with the vertical harmonic model where the ground-state surface is expanded around the excited-state equilibrium geometry (Franck--Condon point) instead of the ground-state minimum energy point. If the potential energy surface were completely harmonic, the two different harmonic approximations should give identical results.
Figure~\ref{fig:lha_vha_spectra} shows that the $12^2$ SVL spectrum obtained from the vertical harmonic model has
 notable differences in peak intensities and positions compared to the adiabatic harmonic spectrum (top left). Although the overall structure remains similar, the adiabatic harmonic spectrum (bottom left, fig~\ref{fig:spectra}) provided better agreement with the experiment than the vertical harmonic one (bottom left, fig~\ref{fig:lha_vha_spectra}).

\begin{figure}
    \centering
    \includegraphics[width=\linewidth]{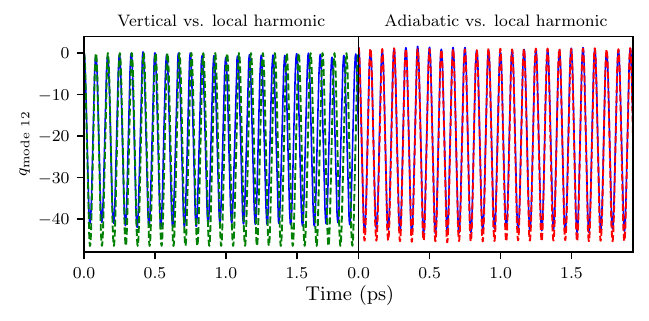}
    \caption{Evolution of the position along mode 12 ($q_\text{mode 12}$) of the center of the wavepacket in the on-the-fly local harmonic dynamics (blue solid line), compared to the vertical (left panel, green dashed line) and adiabatic (right panel, red dashed line) harmonic dynamics.}
    \label{fig:q_mode12}
\end{figure}

To confirm the impact of anharmonicity in anthracene, we also performed the much more expensive on-the-fly ab initio local harmonic calculations.\cite{Zhang_Vanicek:2025} Here, we expand the potential to second order based on DFT evaluations of gradients and Hessians at the current position, rather than using a state-independent global harmonic potential.
The center of the wavepacket then follows a fully anharmonic classical trajectory, while the evolution of the width (matrices $Q_t$ and $P_t$) is determined by the local Hessian at each time step.
The local harmonic approximation produced very similar results to the global adiabatic harmonic approach (top right, fig~\ref{fig:lha_vha_spectra}).
Figure~\ref{fig:q_mode12} compares the position evolution in mode 12 for the three approximations. The adiabatic and local harmonic trajectories are nearly identical with small differences in the amplitude of motion, whereas the vertical harmonic model gives a significantly different trajectory in the amplitude and frequency of oscillation.
This suggests that the (adiabatic) harmonic model suffices for the evaluation of SVL spectra of anthracene.

However, empirical wavenumber scaling is still required for the local harmonic results to achieve good agreement with experimental wavenumbers (bottom right, fig~\ref{fig:lha_vha_spectra}). Although the use of scaling factors are typically attributed to anharmonicity, our results suggest they may be primarily compensating for systematic bias (e.g., due to electronic correlation) in the chosen DFT method.
While outside the scope of the present work focused on the dynamics method, further studies using anharmonic frequency corrections (e.g., with perturbation theory\cite{Barone:2004,Fuse_Bloino:2024}) could help separate electronic structure errors from genuine anharmonic contributions.
The sensitivity of the computed spectra to the different harmonic models in figs~\ref{fig:effects} to~\ref{fig:lha_vha_spectra} reflects how vibronic spectra may reveal detailed information about molecular potential energy surfaces in experiments.

We also note that the additional cost of computing overlaps for different initial states is negligible compared to propagation in the local harmonic case. As a rough comparison, in the global harmonic model, propagating the parameters of the Gaussian for 100,000 steps took approximately 12 CPU minutes (single-core) in total, while computing 25,000 overlaps for a doubly excited level required around 72 CPU minutes on the same computer. Whereas constructing the global harmonic potential for anthracene required only a single electronic structure evaluation (approximately 90 CPU minutes), on-the-fly local harmonic dynamics required performing such a calculation at \emph{each} time step.

Since the overlap expressions for two Hagedorn functions are general for any non-negative multi-index $K$, our method can use the same trajectory to treat not only multiple excitations in a single mode, but also initial states where several vibrational modes are simultaneously excited to any vibrational level (although the cost of computing overlaps increases as more modes are excited to higher levels).
 Here, we have restricted ourselves to the experimentally accessible vibronic levels ($12^{1}6^{1}$, $12^{1}\overline{5}^{1}$, and $12^{1}5^{1}$), which were reported respectively at $1380+385\,$cm$^{-1}$, $1409+385\,$cm$^{-1}$, and $1420+385\,$cm$^{-1}$ in the experimental fluorescence excitation spectrum of anthracene, albeit with very weak intensities (2--3\% of the signal at origin).\cite{Lambert_Zewail:1984} Despite the current lack of experimental data on the SVL fluorescence from these levels for a direct comparison, the Hagedorn approach can serve as a tool for predicting complex vibronic spectra.

\begin{figure}
    \centering
    \includegraphics[width=0.8\linewidth]{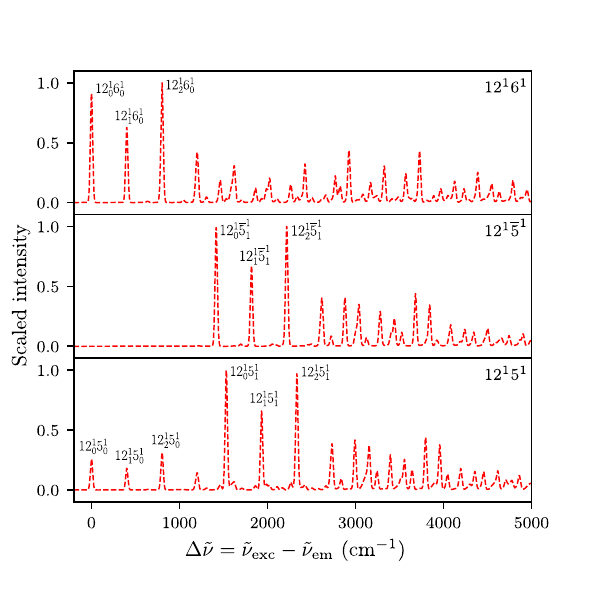}
    \caption{SVL fluorescence spectra of anthracene from initial vibrational levels $12^{1}6^{1}$, $12^{1}\overline{5}^{1}$, and $12^{1}5^{1}$ predicted with Hagedorn wavepacket dynamics in the adiabatic harmonic model; a scaling factor of 0.97 was applied to the wavenumbers.}
    \label{fig:spectrum_predict}
\end{figure}

Figure \ref{fig:spectrum_predict} shows the SVL emission spectra from the levels $12^{1}6^{1}$, $12^{1}\overline{5}^{1}$, and $12^{1}5^{1}$ predicted with the adiabatic harmonic model.
These spectra are broadly similar to the simulated $12^1$ spectrum, but the excitation in an additional mode leads to different intensity patterns.
Compared to the $12^{1}6^{1}$ spectrum, the predicted spectrum from the $12^{1}5^{1}$ level 
shows weaker intensities in the first three peaks corresponding to a progression in mode 12.
In constrast to $12^{1}6^{1}$ and $12^{1} 5^{1}$ spectra,
in the $12^{1}\overline{5}^{1}$ spectrum the three peaks of the mode 12 progression under $1000\,$cm$^{-1}$ (i.e., $12^{1}_{j}\overline{5}^{1}_{0}$ with $j=0,1,2$) disappear, because transitions to $\mathrm{a_g}$ levels (e.g., the ground level and $12_{j}$ levels) in the ground state are forbidden
for initial levels of $\mathrm{b_{1g}}$ symmetry.
The progression in mode 12 is only allowed when coupled with the symmetry-allowed $\overline{5}^{1}_{1}$ transition (e.g., $12_0^1 \overline{5}_1^1$, $12_1^1 \overline{5}_1^1$, and $12_2^1 \overline{5}_1^1$). Otherwise, the features above $2500\,$cm$^{-1}$ (mainly combination bands with other $\mathrm{a_{g}}$ modes, analogous to the ${>}1000\,$cm$^{-1}$ peaks in the $12^1$ spectrum) are similar in the SVL spectra from $12^{1}\overline{5}^{1}$ and $12^{1}5^{1}$ levels.

\section{Conclusions}

To conclude, we have combined Hagedorn wavepacket dynamics with DFT evaluation of the electronic structure in order to simulate SVL spectroscopy in a realistic molecular system. From a single semiclassical trajectory, we were able to efficiently compute SVL spectra of anthracene beyond emission from singly excited vibrational levels and consistent with symmetry considerations and in good agreement with experiments.

Because the vibrational Hagedorn dynamics in the global harmonic model is exact, it is easier to identify and isolate errors.
Our analysis of the results with different harmonic approximations demonstrates the importance of including both mode distortion and Duschinsky rotation in evaluating molecular vibronic spectra, which serve as a sensitive probe for molecular structure and dynamics. 
For anthracene, our results suggest that the electronic structure methods we used underestimated mode mixing, which affected the accuracy of combination band intensities.

The Hagedorn approach can be readily extended to capture mild anharmonic effects through the local harmonic approximation, which requires expensive on-the-fly electronic structure evaluations along the trajectory (though the same propagation can again be used to evaluate spectra from all initial excitations).\cite{Zhang_Vanicek:2025}
Computing vibronic spectra within different global harmonic models can help selecting appropriate ab initio methods and identifying when anharmonicity becomes significant before more sophisticated quantum dynamics methods are applied. 

Even within the global harmonic framework, an extension of our method could be useful
for other experimental techniques involving vibrationally excited states, such as vibrationally promoted electronic
resonance (VIPER) experiments,\cite{VanWilderen_Bredenbeck:2014,
VonCosel_Burghardt:2017, Horz_Burghardt:2023} time-resolved photoelectron spectroscopy,\cite{Yu_Ullrich:2014} and fluorescence-encoded
infrared (FEIR) spectroscopy.\cite{WhaleyMayda_Tokmakoff:2021}
More generally, Hagedorn functions form a complete orthonormal basis and can exactly expand arbitrary polynomials times a Gaussian.\cite{book_Lubich:2008,Lasser_Lubich:2020}  This could enable Hagedorn wavepackets to treat more complex initial states, for example, in non-Condon spectroscopy\cite{VonCosel_Burghardt:2017, Patoz_Vanicek:2018,Begusic_Vanicek:2021a,Gandolfi_Vanicek:2025}. Finally, to account for non-adiabatic effects, one could extend the approach by variationally propagating multiple Hagedorn bases on coupled electronic surfaces.\cite{Adhikari_Billing:1999,Ben-Nun_Martinez:2000,Worth_Burghardt:2008,Bourquin_Hagedorn:2012}

\section*{Data Availability}

In addition to the data provided in the article and the Supporting Information, the electronic structure results used to construct the initial state and the global harmonic models, along with the computed spectra and autocorrelation functions, are available on Zenodo (DOI: 10.5281/zenodo.15693437).

\suppinfo

Expressions to compute the overlaps between Hagedorn functions, optimized $S_0$ and $S_1$ geometries of anthracene with computed and experimental frequencies of selected vibrational modes, additional computational details of the empirical Duschinsky coupling and of the local harmonic calculation, computed and experimental $0^0$ emission spectra, computed spectra in figs~\ref{fig:spectra} and~\ref{fig:lha_vha_spectra} without wavenumber scaling.

\begin{acknowledgement}
The authors acknowledge financial support from the EPFL.
\end{acknowledgement}

\bibliography{hagedorn_svl_anthracene_v20.bib}
\clearpage 
\pagestyle{empty}
{
 \renewcommand{\newpage}{\par\pagebreak}
  \includepdf[pages={-},angle=0,offset=0 0,noautoscale=true]{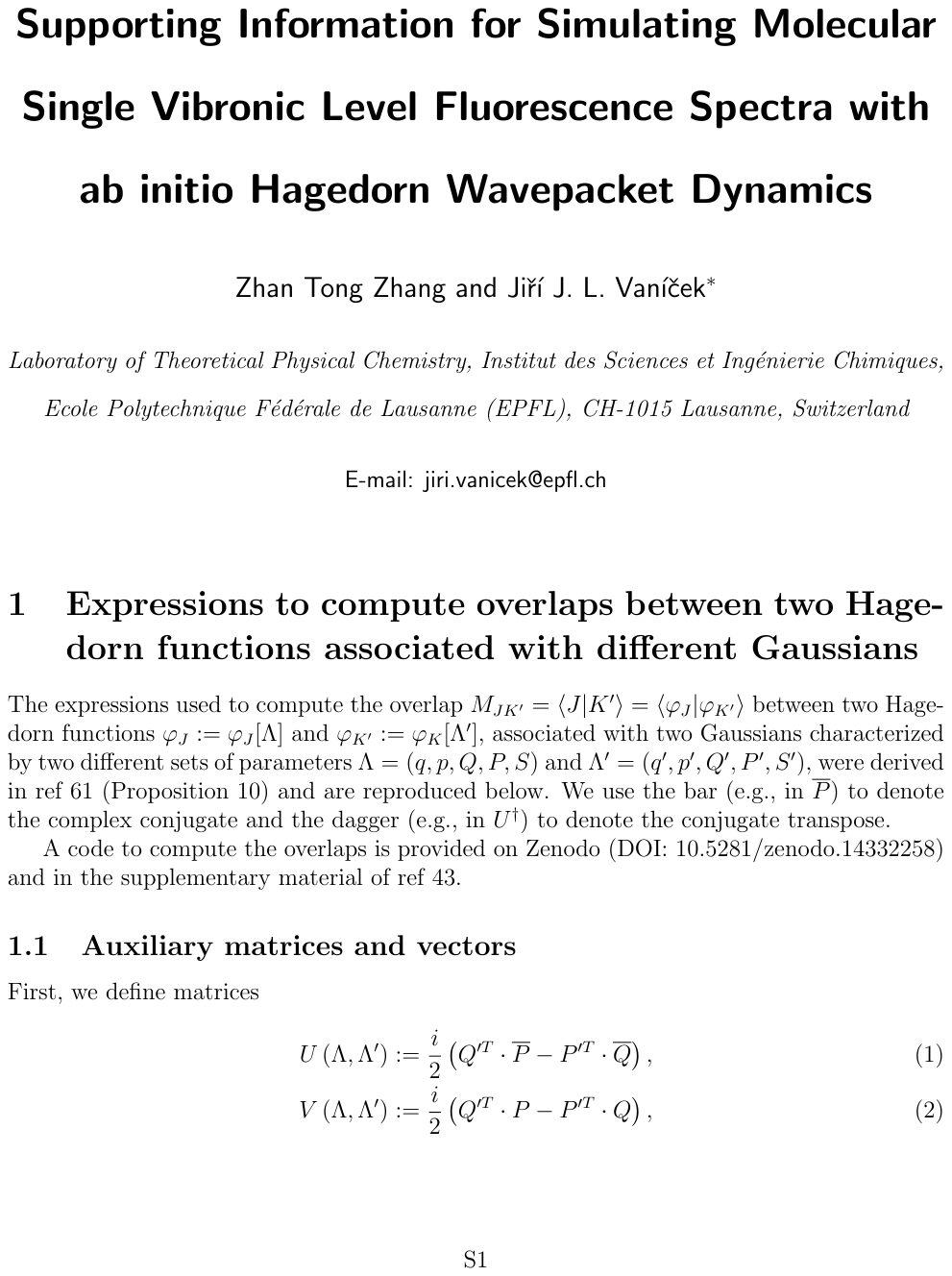}
}
\end{document}